\documentclass[conference]{IEEEtran}

\ifCLASSINFOpdf
\else
\fi

\usepackage[hidelinks]{hyperref}
\usepackage[noadjust]{cite}

\usepackage{amssymb}
\usepackage{amsmath,amsfonts}
\usepackage{multicol}
\usepackage{color,graphicx}
\usepackage{epsfig}
\usepackage{subfigure}
\usepackage{setspace}
\usepackage{cite}
\usepackage{enumerate}
\usepackage{algorithmic}
\usepackage{algorithm}
\usepackage{array}
\usepackage{multirow}
\usepackage{epstopdf}
\usepackage{dblfloatfix}
\usepackage{fixltx2e}
\usepackage{tcolorbox,bbold}
\usepackage{graphicx}
\usepackage{kantlipsum} %<- For dummy text
\usepackage{mwe} %<- For dummy images

\usepackage[noadjust]{cite}
\usepackage[belowskip=-1pt,aboveskip=0pt]{caption}
\usepackage{xcolor}

\DeclareCaptionLabelSeparator{periodspace}{.\quad}

\newcommand{\bc}{\begin{center}}
\newcommand{\ec}{\end{center}}
\newcommand{\bt}{\begin{tabular}}
\newcommand{\et}{\end{tabular}}
\newcommand{\bi}{\begin{itemize}}
\newcommand{\ei}{\end{itemize}}
\newcommand{\be}{\begin{enumerate}}
\newcommand{\ee}{\end{enumerate}}
\newcommand{\beq}{\begin{equation}}
\newcommand{\eeq}{\end{equation}}
\newcommand{\comments}[1]{}

\begin{document}
%Here goes the title
\title{Enterprise API Security and GDPR Compliance: Design and Implementation Perspective }
%\markboth{Innopolis University}{}
\author{\IEEEauthorblockN{Fatima Hussain\IEEEauthorrefmark{1}, Rasheed Hussain\IEEEauthorrefmark{2},
Brett Noye\IEEEauthorrefmark{1},  and Salah Sharieh \IEEEauthorrefmark{1}}
 \IEEEauthorrefmark{1} Royal Bank of Canada (RBC), Toronto, Canada \\
\IEEEauthorblockA{\IEEEauthorrefmark{2} Networks and Blockchain Lab, Institute of Information Security and
Cyber-Physical Systems, Innopolis University, Russia\\}}

%Email: \IEEEauthorrefmark{2}\{r.hussain@innopolis.ru 
%\IEEEauthorrefmark{1}fatima.hussain@rbc.com}}

\maketitle
%Main body starts
%\input{sections/0_abstract}
\begin{abstract}
With the advancements in the enterprise-level busi-
ness development, the demand for new applications and services
is overwhelming. For the development and delivery of such ap-
plications and services, enterprise businesses rely on Application
Programming Interfaces (APIs). In essence, API is a double-
edged sword. On one hand, API provides ease of expanding
the business through sharing value and utility, but on another
hand it raises security and privacy issues. Since the applications
usually use APIs to retrieve important data, therefore it
is extremely important to make sure that an effective access
control and security mechanism are in place , and the data
does not fall into wrong hands. 
In this article, we discuss the current state of the
enterprise API security and the role of Machine Learning (ML)
in API security. We also discuss the General Data Protection
Regulation (GDPR) compliance and its effect on the API security.

% We furthermore, discuss the future challenges of the ML-based API security.
 
 \end{abstract}

\begin{IEEEkeywords}
 API Security and GDPR, Enterprise API Security, Automated API Security
\end{IEEEkeywords}

\section{Introduction}
 With the advancements in communication technologies, there is myriad of applications and services that target consumers in different sectors ranging from finance, health, agriculture, smart industries, smart environment, and human well-being \cite{fernando2018}. Internet of Things (IoT) is the best example of such applications and services realized through the interconnection of smart object for different purposes such as, but not limited to, controlling home appliances remotely, monitoring patient's health condition, monitoring agricultural land, operating in hostile environments, and so on. In most of the cases, applications and services in these domains have more than one stakeholders. Furthermore, these services are also shared across different platforms with different consumers, vendors, and other related entities. Therefore, management of these services and applications, and expanding them across different domains and consumers, the traditional off-the-shelf software development solutions will not scale well. Therefore, we need a unified mechanism to make the applications and services (both macro- and micro-services depending on the application) easy to access, secure, able to export, and meet the heterogeneous consumer demands. To this end, Application Programming Interface (API) is a mechanism that makes it easy, affordable, and scalable for the services to distribute across different domains. More precisely, API is a set of protocols, functions, mechanisms, tools, definitions, and attributes to share and develop new services across different domains and expand the existing services. APIs enable service integration, application development, and communication among different services and product without need for developing new infrastructure for each service and product. For instance, in case of a financial institution; banking chatbots, reduction in cost due to decoupling of platforms and rejoining through APIs, fast agility change, enhanced operational efficiency  and availability of new distribution channels are few benefits of using APIs for banking.  Also, banks  use APIs internally to improve information flow between various legacy systems. %Banks are adopting Open Banking concept by proactively exposing their systems/data available to third parties for generating new revenue streams.
With rise of IoT, world around us is more connected, and  APIs  has emerged as an integral business strategy across various  industries. APIhound reports more than 50,000 registered API till date. Also,  private  APIs exceed in number compared to number of public APIs.  This leads to challenges of security and privacy as lots of sensitive data being passed over the web through APIs.

In this paper , we discuss API security, what does it mean and  what is being done traditionally? afterwards we talk about  inadequacy of the current API security measures and drive discussion towards AI or ML driven API security. Without loss of generality, we discuss the importance of user data privacy and security,  and how evolution of General Data Protection Regulation (GDPR) have changed the paradigm of user privacy. At the end we discuss that how GDPR is going to affect the AI or ML driven API security. 

\subsubsection{Summary of contributions}
The main contributions of this paper are summarized below: 

In this paper, we,
\begin{itemize}
\item Discuss security related issues in ML driven APIs,
\item Briefly discuss GDPR compliance requirements related to security and privacy,
\item We discuss the role of GDPR on ML-driven API security in enterprise environments.
\item Challenges faced in terms of design, customer satisfaction and data transparency by ML-API for being GDPR compliance, 

\end{itemize}

The remainder of the paper is organized as follows:  In Section II, we discuss APIs, API vulnerabilities and API security models.  In Section III, we discuss GDPR and its rules contradicting with ML driven decisions. API security and GDPR compliance is presented in  Section IV.  Finally, paper is concluded in Section V.

\section{API Security}
In this section, we discuss the API security in detail. We focus on the vulnerabilities of API, the traditional API security model and the role of ML in API security.

\subsection {APIs and Their Vulnerabilities}
APIs are functionally classified into two categories. The APIs that are ``used to perform an action'' or APIs that ``provide an access to any object". In the former type of APIs, an application invokes the API and requests the original software to perform an action (which is made available through the invoked API). On the other hand, in the latter type APIs, an application wants to get access to an object through the API.  Broadly speaking, there are three types of APIs: \cite{Fatima-API}

\begin{itemize} 

\item {\it Private API}: 

Private APIs are usually intended to be used solely by the firm making the software. Companies develop private APIs for internal software development and enhancing/providing scalability, modularity, security, access, for expanding various services. 

\item {\it Partner API}: 

APIs developed for usage among partners are known as partner APIs.  For instance, a firm develops a software package for sales and marketing functions, where another partner firm  has a software for accounting. These two softwares can be connected and and this integration is bridged through APIs. However, it requires efficient access control and authorization mechanisms, along with rules and policies of the firms involved in software development and service delivery. 

\item {\it Public API}:

Public APIs are intended to be used by anyone who wants to access the software. These APIs have limited capabilities and can be  a potential security threat to the back end system. More precisely, the attackers could launch attacks camouflaged into the functions and services provided by the APIs. 
\end{itemize}

\subsubsection{Vulnerabilities}
In contrast to the typical websites, the APIs open up wide access for the clients and also lure  potential hackers into the back-end systems. The potential attack surface is significantly increased by using APIs as granularity boundary is moved from secure internal tiers to the user devices (through client application). Therefore, we need security and protection mechanisms from the new class of risks as a result of using APIs, in addition to the traditional threats (carried out from the Web). 

In a conventional web scenarios, there are only few ways of data rendering,  as some data is sent to remote systems (servers) and it mainly depend upon the capabilities of URLs and forms.  While, APIs expose more of the HTTPS protocol and open up data parameterization.  As a result, when the data sent to APIs (if not handled with care), increase the potential attack surface through parameter attacks such as URL-based attacks, query parameters, HTTP header and/or post content and so on. %potential attack surface is stretched and vulnerable to parameter attacks including URL, query parameters, HTTP headers and/or post content, when data is sent to APIs. Few parameter attacks are:
 In the following, we outline some potential parameters attacks on APIs.

\begin {itemize} 
\item Script insertions: It refers to the family of attacks that  exploit the systems that interpret the submitted parameter content as a script (e.g. when a snippet of JavaScript is submitted into a posting on a Web forum).

\item SQL injections: It refers to an attack through query languages where parameters that are designed to load a certain input into a database query, are manipulated to change the intent of an underlying SQL template.

\item Bounds or buffer overflow attacks: It provides data beyond the expected types or range, and leads to system crashes and also offer access to memory spaces.
\end {itemize} 

In the same spirit, there are few access related attacks specific to APIs listed below. Conventional access control mechanism such as; user name passwords, openID, JWT tokens etc. are powerful but leaves security gaps in API deployment. These techniques requires complementary security capabilities to address threats such as:
\begin{itemize}
\item API-specific DDoS attacks: These attacks overload critical API services (login and session management)  and disrupt access to these services, by sending large amounts of traffic from multiple sources. 

\item Login attacks: These attacks includes, credential stuffing (testing lists of previously breached credentials against a target API to try to gain access), use of stolen credentials or tokens,  or fuzzing (feeding large amounts of random data into a program to discover vulnerabilities).

\item Application \& data attacks: These attacks include data theft, data deletion or manipulation, code injection,  and  application disruption. 
\end {itemize}

We can categorize API security vulnerabilities systematically, by target areas. These categorization also takes into account where and how various attacks on API potentially breech the security. For instance:

\begin{itemize}
\item Network, operating system, and driver issues are related to operating system and network components such as buffer overflows, flooding with sockets, and Denial of Service (DOS) attacks etc.
\item Application layer issues are related to hosting application, server and related services such as message parsing, session hijacking or security mis-configurations.
\item API component functional issues are related to actual APIs such as injection attacks, sensitive data exposure, incomplete access control and so on.
\end {itemize}

\subsection{Traditional API Security Model}
\begin{figure}
\centering
\includegraphics[scale=0.4]{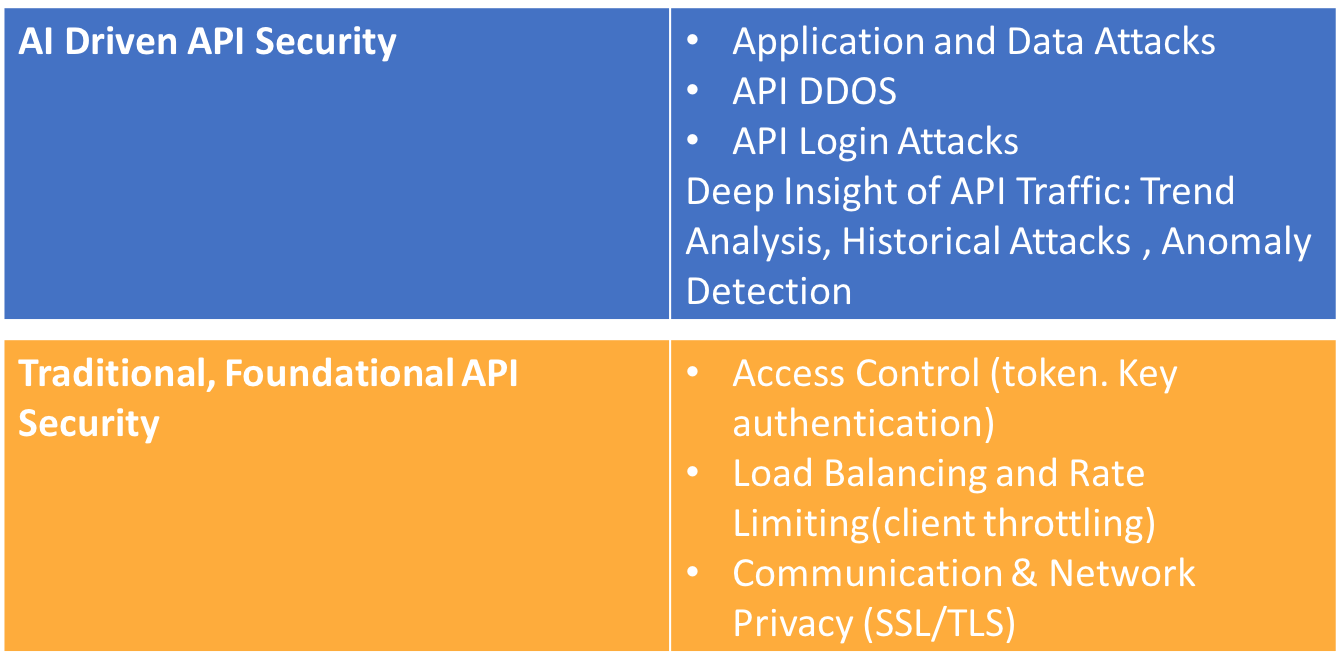}
\caption{API Security Model }
\label{model}
\end{figure}

Traditional API security model incorporates tasks related to  authentication, throttling, and communication security as shown in the Figure \ref{model}.  These are powerful tools but are not considered comprehensive solution for addressing specialized API threats; such as API-specific denial of service, application and data and log-in attacks \cite{Fatima-API}. Therefore, a comprehensive API security solution requires anomaly detection capability as well as basic security capabilities. On the contrary, AI-enabled API security keep track of historical traffic trend along security with existing foundational security features  and detect malicious behavior as a first line of defense. Traditional security measures provided by Content Distribution Networks (CDNs), Web Application Firewall (WAFs)   and API Gateways can be easily bypassed by finely tuned attacks on APIs.

API implementations are based on either REpresentational State Transfer (REST) or Simple Object Access Protocol (SOAP),  and are secured in different ways as discussed in the following subsections. Generally speaking, SOAP APIs  have more comprehensive security measures and are recommended for handling sensitive data.

\subsubsection{Access control management}
By granting or rejecting an access to APIs is the first line of defence for the internal resources. Controlling the amount of data released to the cyber-world is possible by limiting access to specific endpoints or data for individual clients. %An internal developer, for example, will likely need wider access to a set of data to accomplish their job than a third-party developer. 
Access management is typically performed using “key”  to identify applications calling the APIs as well as the end users. This key has access to specific endpoints and has access privileges for certain data limit. 

To this end, Open Authorization (OAuth) and OpenID are used for user authentication and authorization for the web services. OAuth is the open standard for access management and it enables users to have access to API resources without sharing passwords. OAuth is complimented with another standard, OpenID Connect (OIDC). This is an identity layer on top of the  OAuth framework, and it authenticates users by obtaining the basic profile information. 

\subsubsection{Communication security}
Transport Layer Security (TLS)/ Secure Socket Layer (SSL) is used for the communication security of any web service. TLS standard is used to establish secure connection between two endpoints (client and server) and it makes sure that the data sent between them is encrypted and unaltered.  REST APIs use HTTP and is supported by TLS encryption. It also uses JavaScript Object Notation (JSON) (a file format used to transfer data securely and efficiently over web browsers). By using HTTP and JSON, REST APIs do not need to store or repackage the data, and are considered faster and easily manageable than the SOAP APIs.
SOAP APIs use built-in protocols such as Web Services Security (WS Security) %and support  two standardization bodies, the Organization for the Advancement of Structured Information Standards (OASIS)  and the World Wide Web Consortium (W3C). 
and use a combination of XML encryption, XML signatures, and SAML tokens to verify authentication and authorization.

\subsubsection{Client throttling} 
Client throttling enables the access limits for APIs, i.e, how often an API can be called? and also track its usage over certain time period. Carefully crafted throttling rules can protect APIs from spikes and DoS attacks. For instance, more calls to an API indicate that it might being abused, or it might be a programming mistake and API is being called in an endless loop. This  information is very useful for identifying and preventing various access related issues.

\subsubsection{API gateways security} 

The API gateway is considered as the core infrastructure unit that enforces and manages API security. Enforcing security through API gateway is comparatively new concept and serves better for API security unlike traditional security measures. API security management performs  message analysis, access tokens and authorization parameters grants, and therefore API gateway checks authorization of users followed by message parameter's and content checks (sent by authorized users). It also ensures that the client data is not written when usage logs are maintained. Hence, API gateway  acts like a traffic police and ensures that only legitimate users are allowed access to APIs and rest of them are blocked. It also encrypts or redacts( censor or obscure (part of a text) for legal or security purpose) confidential information, control, and analyze the APIs usage. Essentially, with the help of an API gateway, we are moving security from the application into the organizational infrastructure.

\subsection{Limitations of Traditional API Security Model} 

 As discussed already, API has the pivotal role in the realization of the Programmable Web. Till the start of 2018, a steep growth has been observed in the web APIs \cite{Ping}.%\footnote{\url{https://www.pingidentity.com/en/company/blog/posts/2018/artificial-intelligence-machine-learning-a-new-approach-to-api-Security.html}}. 
 As the business grows, the inter-operability among different components of the same business or with the partner businesses is crucial to the growth and expansion of the business and thus needs paramount of attention. APIs are the only viable way to address this issue and the security of these APIs is fundamentally essential. The traditional security mechanisms such as OAuth and others (as mentioned already) focus on only the visible aspects of the security such as authentication, access control, and authorization. However, with the growth in API development and the emergence of new APIs, it also increase the risk of the exposure of sensitive data beyond the business boundaries. The traditional approach of {\it "limiting access to the API"} instead of mitigating the attacks, has not been so encouraging. This argument is based on the attacks against sophisticated APIs so far. Furthermore, every new API brings a long a new attack vector associated to it. Therefore, it is quite hard to address the security attacks on APIs through a singular traditional approach such as access control. In fact, there are targeted attacks that disrupt the normal functionality of the APIs. For instance, (Distributed) Denial of Service (DDoS) attacks against log-in services are still possible against APIs incorporating strong access control mechanisms%\footnote{\url{https://www.pingidentity.com/en/platform/api-security.html}}. 
 Similarly, the stolen or shared credentials may also result in catastrophic attacks on the API. Additionally, the traditional injection, data stealing, and manipulation attacks are still possible in APIs. The traditional APIs may not be able to mitigate these types of attacks. Therefore, more versatile, variable, context-aware, and intelligent security mechanism is needed for API security. In the following, we describe the need for Artificial Intelligence (AI) and Machine Learning (ML)-based API security.  

\subsection{Machine Learning and API Security} 
With the evolution of Cyber-Physical Systems (CPSs) and the diversification of threat model, it is essential to make the smart API security as an integral part of the API operations. Machine Learning (ML) can be used not only to identify the malicious intent in transactions of data across platforms but it also helps evolving the security practices in the wake of current security practices. To date, ML has been widely used in the security of systems and networks, for instance, context-aware authentication, authorization, intrusion detection, malware analysis, and so on \cite{Xin2018,UCCI2019}.  %perfect tool not only for identifying malicious intent in transaction data but also helpful in evolving security practices as threats evolve.

 In the context of API security, ML is primarily leveraged to learn patterns of normal behavior incorporating the contextual information for each API. The identified patterns are then used to identify and block potential cyber attacks on the APIs. Continuous learning capabilities are added to the system and APIs through which anomalous behavior can be identified, even without written policies or prior knowledge of common attack patterns ( zero-day attacks). In short, ML can extend the API security beyond access control and communication security , and help filling the security gaps such as addressing new cyber threats, identifying the behavior of the past attacks, making predictions on the basis of existing patterns to manage the API security, and so on.
 
Detailed discussion about ML-based API security, available techniques and platforms can be found in  \cite{Fatima-API}.

\section{General Data Protection Rule (GDPR)}

\begin{figure*}
\centering
\includegraphics[scale=0.6]{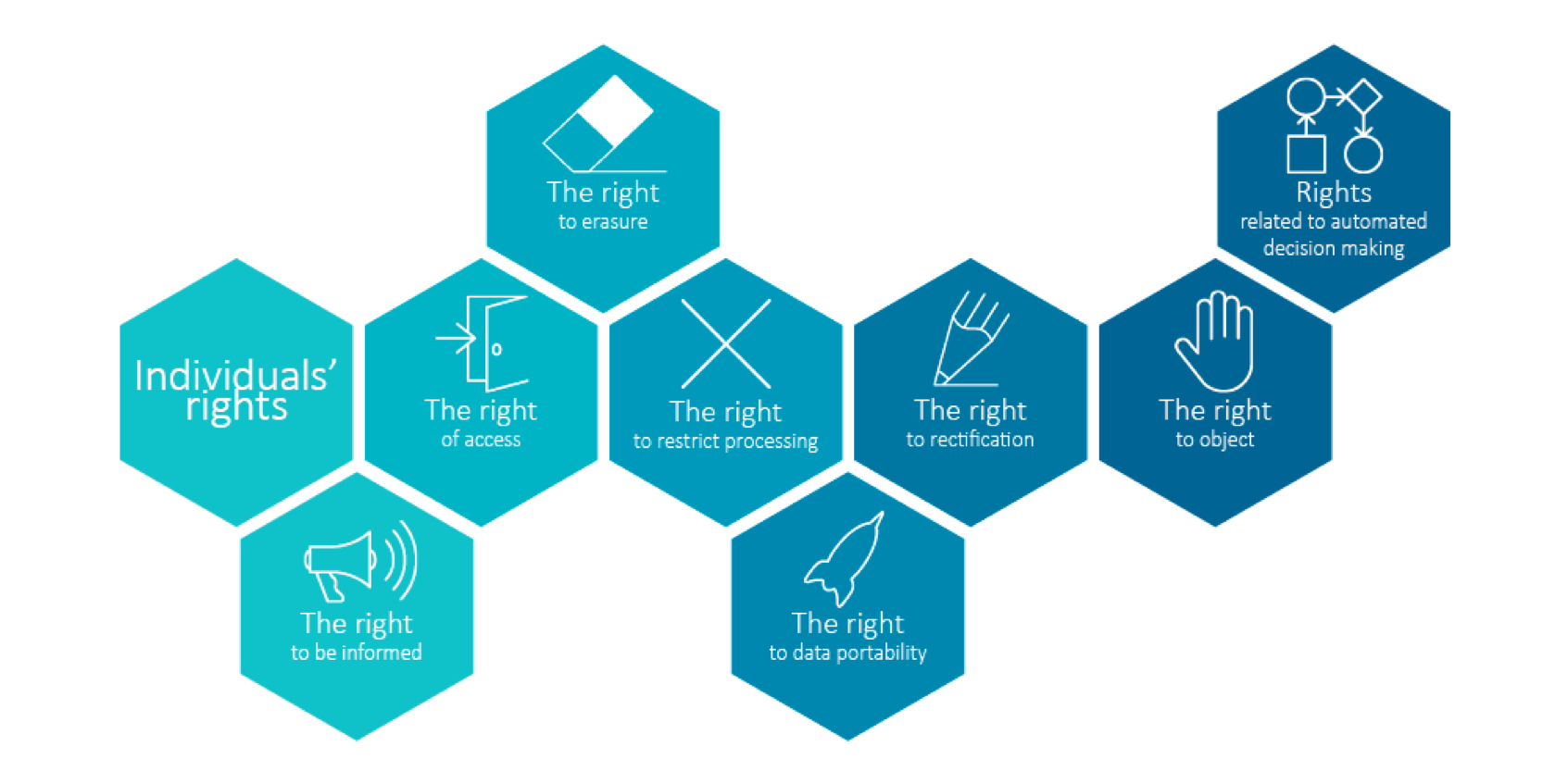}
\caption{General Data Protection Regulation (GDPR)\cite{GDPR-C}}
\label{GDPR}
\end{figure*}
%\footnotetext{https://hackernoon.com/how-to-make-your-product-gdpr-compliant-396a6c0336c2}

European Union Parliament approved a revolutionary regulation on the personal data protection in April 2016 and it is functional since May 2015. The enforcement of GDPR can be seen as reshaping the personal data protection against misuse by the legitimate entities, such as the firms that collect the customer data. 
GDPR emphasizes on personal data protection, transparency and data ownership rights for individual users. In addition to this, it also gives right and access to users, how they wish to get their data treated, as shown in Figure \ref{GDPR}. Personal data in GDPR is defined as any information that can identify any individual directly or indirectly, and "Personal Data Processing" is defined as the set of automated/manual operations, performed on the personal data \cite{Wong}.  These operations include data collection, recording, organization, structuring, storage, adaptation or alteration, retrieval, consultation, use, disclosure by transmission, dissemination, alignment or combination, restriction, erasure or destruction \cite{GDPR}etc.  GDPR applies to companies in European Union, as well as any company across the globe that processes the personal data of any EU's individual. This processing can be either (i) goods/services offerings to data subjects in the EU, or (ii) their behaviour monitoring.

From enterprise perspective, GDPR requires:  
\begin{itemize}
    \item Transparency: Clear policies must be defined  for data protection, data processing and data portability of the customer related  information. 
    \item Access control: Enterprises must possess proper security tools as well as processes for protection of the customers’ private data.
    \item Personal privacy and right to be forgotten: A customer older than 16 years of age has full right to dictate what type of data an enterprise can collect about that customer. Furthermore, the customer has full right to demand his/her data to be deleted after usage.  
    
\end{itemize}

\subsection{Contradiction of GDPR Rules with ML}
GDPR grants three important rights to the owner (data subject) of personal data: {\it right of non-discrimination}, {\it right to explanation}, and the {\it right to be forgotten}. In the following, we discuss these rights with respect to ML work flows, i.e, feature engineering and data wrangling. We also discuss  ML modeling development, model deployment, and model  management. 

\subsubsection{Feature engineering and non-discrimination right}

Personal data processing rules are the pillars of GDPR, i.e, revealing racial/ethnic origin, political opinions, religious  beliefs,   bio-metric data used for identification purposes, health data, data related to sexual orientation, and the processing of genetic data are protected by GDPR.
For instance,  these  data points are incredibly valuable in genetic research, and being used for predictive modelling in the different domains. However, explicit consent is required by the data subjects for opting in for such model training as well as for on going model retraining for improving the model accuracy.   

\subsubsection{Modelling, prediction and right to explanation}

Clients, customers, and/or users have the right to understand the processing logic and reasons for any potential decisions made for or on behalf of them. Therefore,  the Processors (enterprise) are bound  to provide meaningful information about the decision logic and justification of any prediction  and envisaged consequences of this processing for the data owner. However, this is very difficult for authorities to decide , how deep  this explanation right should go? Is it essential to explain all  the data transformation-related details to the data owner? How difficult it is to interpret the entire  predictive modeling processes, specifically when multiple algorithms are involved in a work flow?

\subsubsection{Model retaining/updates and Right to be Forgotten}

This right permits data owners to dictate processors (enterprise) to erase all personal data associated with him/ her. Apparently, it seems very  straightforward to delete corresponding account and related data.  However, it poses lots of technical challenges for ML models, if retraining and regular update of ML model is required which in turns require availability of one's data.  Does data owners ts have the right to authenticate first weather their data is used to retrain the  predictive model or not?  How ML model will be retrained without data?   Where the line should be drawn in terms of amount of data to be retained and to be forgotten. 

\subsubsection{Global privacy and data localization }

It deals with all the issues related to the storage and
transfer of personal data  across the countries or regions.  GDPR defines this personal data as "any data revealing racial/ethnic
origin, health related data, religious/philosophical beliefs, sexual orientation data, and genetic/biometric  data". These perceptions of human abilities and personal interests make human profiles. In todays's world of "Internet of Things",  profiling is  carried out by machines using ML algorithms. ML is used for data mining of the available personal data to obtain important information from the available commercial databases such as maintenance records, loan applications, financial transactions, and medical records. These records are fed to ML algorithm by the data controller or processor (third party cloud) or by both. These personal data processing or profiling might requires real-time data processing (depending on the type of application), which  might happen locally at  data controller as well as sent to the cloud for dynamic training of the
algorithm. In this situation, data portability with individual's consent becomes complex and time consuming.

GDPR  also argues on the decisions made as a result of profiling. It does not limit profiling but requires that a  decision made on such profiling is made in a way that it does not have legal or any other significant effects on the individual. 

Also, with the increasing trend in trans-border personal data flows in today's data-driven economy, it is becoming very important as well as difficult for any jurisdiction’s capacity to enforce personal data protection laws beyond its territory. GDPR restricts the rights of personal data, its  portability, and  processing solely to the individual itself. A controller or processor is  any "person, public authority, agency or other body which processes personal data on behalf of controller'', and  can transfer any  personal data to a third country or any international organization after ensuring and providing appropriate safeguards. It is responsibility of controller and processor to make sure that enforceable  rights and effective legal remedies for data subjects are made available.

In case of a security breach of the data  involving personal data, the controller (can alone or jointly with others, determines the purposes and means of processing personal data) should notify Data Protection Authority (DPA) within a reasonable duration of time. Entire obligation is on data processor and controller to keep record of all types of data storage/processing activities. 

\section{API Security and Compliance with GDPR}
GDPR has a significant effect on ML-enabled security as GDPR imposes restriction on the use of automated decision making including profiling (Article 22) \cite{GDPR}. This leads to an impression that  further development of ML-enabled decisions are hampered. Specifically, in this era of Cyber-Physical System and big data in which  automated decisions and predictive analysis are done in almost every walks of life, GDPR compliance with ML enabled solutions seems unacceptable and un-adaptable.  Nevertheless, it is easier to narrow down and avoid decisions that directly affect individual, however, it is unclear to signify the type of harmful profiling. For instance,  advertisements  sent by Google and Facebook have negative effect or not?  In fact detailed clarification and classification are required to distinguish among various automated decisions, essentially being procedural or substantive, rule-based or law-based \cite{Brkan}.

\subsection{Does GDPR affect the ML-driven API Security?}
GDPR will significantly affect the ML-driven security solutions and will tie down the digital development specifically in EU and to some extend in rest of the world. Consumers routinely interacting with ML enabled services such as; personal assistants chatbots, robo-advisors( providing automated financial advice) and applications using streaming services (movie recommendations)  will be significantly affected. 

GDPR restrictions will increase the cost of ML-driven solutions directly or indirectly. For instance, requirement to explain details of algorithmic decisions to human is not only complex but also time consuming that demands particular skills. The right of data portability does not directly affect ML- driven services, but it increase the cost indirectly. As it restricts the companies for creating and maintaining  large and complex data sets in reusable formats \cite{Europa}. Also, there is a trade-off between algorithmic transparency and
accuracy. Therefore more transparent and less accurate algorithms are developed to explain algorithmic decisions to consumers, and it might lead to unfair decision making. 
Similarly,  prohibition on solely-automated decisions might lead to humans making unfair and un-reasonable decisions.  This will also prohibit use of rational
algorithms , which are adaptable to modification in data and can be adjusted over time to account for unintended biases \cite{Zarsky}. 

In addition to above, right of data erasure (given to data subject) without any undue delay  (Article 17(1) of the GDPR), can be problematic for ML-driven services as some ML algorithms need to keep the data used in the training. By removing this data 
algorithm’s effectiveness can be impacted, or even it can break the necessary flow of it. As these algorithms tend to use this data by generating new
rules for data (future) processing to improve themselves.

\subsection{Data and Computational Transparency}

Automated decision making is prohibited in GDPR, and it is defined as the decisions made without human intervention. As the personal data of an individual is used in any decision making, it is needed to be transparent to users how and why any decisions are made (Article 13,14)?  However, many challenges are associated with making this transparency to work.   In the following, we will discuss few of the associated challenges:

\subsubsection{Technical challenges} 

According to GDPR, a controller using user data for some automated decision making is obliged to provide meaningful information about logic used in making such decisions. This enables users to express their opinion about these decisions and also have right to challenge them, if required. Questions such as what exactly is needed to be revealed to the owner?', how algorithms (used for decision making)  can be explained?, and how the complexity of an algorithm can be simplified to be explained to the owner?. However, technical obstacles are faced for explaining the algorithms. These obstacles vary with the variation in complexity of algorithms and learning speed. For instance, simple tree based algorithms are easy to explain as compared to neural networks which are almost impossible to explain. Neural networks, ML and deep learning are considered as "Black Box"  and it is very difficult to explain or identify potential point of failure. As these complex algorithms are opaque even for developers and it becomes very challenging to educate non technical users.

\subsubsection{Intellectual property} 

In addition to technical and user awareness challenges, state secrecy and intellectual property issues cannot be overlooked. Algorithmic transparency can lead to the exposure of intellectual property to the public which can not only jeopardize the privacy of the computational  secret  but also can endanger the policies (reasons behind decisions) of the authorities. For instance, tax authority will never like to reveal algorithms used to select tax payers for secondary/detailed review. 
Similarly, a financial institution  will not disclose the ML model used for mortgage percentage or interest rates.  

%\subsubsection{IP Related Issues}

 IP-related issues are other pressing issues that hinder the data transparency. According to Trade- Related Aspects of Intellectual Property Rights
(TRIPS) agreement  and  World Intellectual Property Organization (WIPO) copyright treaty, software programs are protected  with Copy Write act.  In this situation, concrete measures should be taken to keep the algorithm transparency as well as maintaining Copy write acts. This means that GDPR's required transparency can be achieved by only sharing "logic behind decisions", and not the algorithm itself.

\subsection{Data Localization and Storage} 
Traditionally, three main types of data localization approaches exists; the jurisdiction-to-jurisdiction, organization-to-organization,
and the data localization approaches \cite{Selby}. Jurisdiction-to-jurisdiction approach governs trans-border data flows based on adequate and equivalent national data protection laws. While organization-to-organization approach put this responsibility on individual data controllers for meeting data protection' standards (when respective data is processed offshore).  Lastly, data localization approach depends on public policy efforts to store personal data within a particular jurisdiction’s boundaries. This is the new trend but underlying meanings and intentions are complex and are associated with specific type of the data such as financial, health or medical records etc. GDPR is in accordance to third approach of data localization.

In light of GDPR and data localization,  major changes are required for the entire ecosystem. New cloud infrastructures are required across EU as well across the globe to accommodate GDPR laws. Major paradigm shift is expected in the data storage systems, and in-house storage. Furthermore, in-house processing and storage  of personal data seems more appropriate solution as compared to public cloud. ((( do you mean appropriate for processing data in private cloud or storage? please read the sentence again and correct it. Because it is very challenging to provide  transparency (ability to directly understand where data is being stored and how data processing is performed) of personal data in public clouds.)))
To keep GDPR compliance, an organization can only transfer data to external processor by ensuring adequate levels of data protection and privacy.  If there are security doubts about any particular destination, controller, or processor, the data cannot travel there \cite{ML-GDPR}. Considering the cost of non-compliance with GDPR, most of the multinational companies will move to in-house processing units.  

Also, as per GDPR, data  processing is restricted to the consent of the data subject. It is also worth mentioning that the corporate customer and employee data might be  dispersed as structured or unstructured data across the cloud, on-premises, or on local/distributed file systems. Therefore, remote controlling (deletion, processing, transfer with subject's consent or intention) of personal data,  stored in the file systems and proprietary cloud might not be straightforward.  It will be very challenging task to obtain an accurate and comprehensive view of the personal data across an enterprise and capture of data across systems.

%\begin{figure}[h]
%\centering
%\includegraphics[scale=0.43]{sections/chart.jpeg}
%\caption{Biggest DDoS attacks in the past 4 years \cite{URL1},\cite{URL2},\cite{URL3}  }
%\label{fig:scr1}
%\end{figure}

\section{Conclusion} 
GDPR strictly advocates for "privacy by design", i.e., data protection computational explainability should be included during the system development  rather than adding it later. Similarly, businesses should practice privacy-preserving analytic methods, business models and techniques such as differential privacy, homomorphic encryption, and federated learning. Therefore, transparency, interpretability, and explainability should be considered by data professionals, ML experts and business personals.

For the existing ML-enabled API security solutions, there are requirements of other softwares or applications to explain an automated decision(s) made by  intelligent software (or machine), to meet the  GDPR requirements. One such effort is Quantitative
Input Influence (QII) \cite{Data}, which is used to reach the transparency of an algorithm.  QII is developed to clarify  and explain ML algorithm (and related factors), that were used in any automated decision making.

In a nutshell, it is imperative to investigate the effect of ML-enabled API security mechanisms on the GDPR. 

\bibliographystyle{IEEEtran}
\bibliography{main}
\end{document}